\documentclass[11pt, executivepaper]{article}
\usepackage[utf8]{inputenc}
\usepackage[T1]{fontenc}
\usepackage{natbib}
\usepackage{amsmath}
\usepackage{mathtools}
\usepackage{xcolor}
\usepackage{amsfonts}
\usepackage{graphicx}
\usepackage{enumitem}
\usepackage{geometry}
\usepackage{hyperref}
\hypersetup{colorlinks= true, allcolors=blue}
\setcitestyle{aysep={}}
\begin{document}

\title{\textbf{Unexpected Quantum Indeterminacy}}

\author{Andrea Oldofredi\thanks{Contact Information: University of Lisbon, Centre of Philosophy, Alameda da Universidade 1600-214, Lisbon, Portugal. E-mail: aoldofredi@letras.ulisboa.pt}}

\maketitle

\begin{abstract}
Recent philosophical discussions about metaphysical indeterminacy have been substantiated with the idea that quantum mechanics, one of the most successful physical theories in the history of science, provides explicit instances of worldly indefiniteness. Against this background, several philosophers underline that there are alternative formulations of quantum theory in which such indeterminacy has no room and plays no role. A typical example is Bohmian mechanics in virtue of its clear particle ontology. Contrary to these latter claims, this paper aims at showing that different pilot-wave theories do in fact instantiate diverse forms of metaphysical indeterminacy. Namely, I argue that there are various questions about worldly states of affairs that cannot be determined by looking exclusively at their ontologies and dynamical laws. Moreover, it will be claimed that Bohmian mechanics generates a new form of \emph{modal} indeterminacy. Finally, it will be concluded that ontological clarity and indeterminacy are not mutually exclusive, i.e., the two can coexist in the same theory.
\vspace{4mm}

\end{abstract}
\vspace{5mm}

\begin{center}
\emph{Accepted for publication in the European Journal for Philosophy of Science}
\vspace{5mm}

\emph{The full text of the paper can be found \href{https://link.springer.com/article/10.1007/s13194-024-00574-9}{here}, DOI: 10.1007/s13194-024-00574-9}
\end{center}
\vspace{5mm}

\clearpage

\section{Introduction}
\label{Intro}

Those who believe in the existence of Metaphysical Indeterminacy (MI) endorse the idea according to which the vagueness or indefiniteness we see in the world---e.g., vague objects as mountains, clouds, or the openness of the future---is neither due to semantic nor epistemic limitations. Thus, it is not the case that our language is unable to capture a certain fact or feature of the world, nor that we lack some knowledge or information about it. On the contrary, MI affirms that the world \emph{per se} inherently hosts indeterminacy (cf.\ \cite{Torza:2023} for an overview).\footnote{For the sake of simplicity, in this paper I will use ``indeterminacy'' and ``indefiniteness'' as synonyms. Although not technically precise, this will not cause issues for the proposed arguments. For details cf.\ \cite{Torza:2023} p.\ 1.}

Remarkably, philosophical discussions about MI have been substantiated with the idea that Quantum Mechanics (QM), one of the most successful physical theories in the history of science, provides explicit examples of worldly indeterminacy. To this regard, it is worth noting that Quantum Indeterminacy (QI), the kind of indefiniteness present in quantum theory, has been investigated since the beginning of QM. It has been suggested, in fact, that already \cite{Schrodinger:1935} raised the question about whether quantum observables have vague or indefinite values (cf.\ \cite{Calosi:2021b}). Similarly, Dirac explicitly referred to indeterminacy in his seminal textbook ``\emph{The Principles of Quantum Mechanics}''. Discussing the notion of superposition of states---in the section titled ``Superposition and indeterminacy''---he wrote:
\begin{quote}
[w]hen a state is formed by the superposition of two other states it will have properties that are in some vague way intermediate between those of the two original states
\end{quote}
\noindent and more eloquently
\begin{quote}
\emph{the superposition that occurs in quantum mechanics is of an essentially different nature from any occurring in the classical theory}, as is shown by the fact that the quantum superposition principle demands indeterminacy in the results of observations in order to be capable of a sensible physical interpretation (\cite{Dirac:1947}, p.\ 13 and p.\ 14 respectively; emphasis in the original).
\end{quote}

In recent years, in addition, QI has become an established field of research for both philosophers of physics and analytic metaphysicians. Indeed, the philosophical analysis of QM individuated different sources of indeterminacy within the theory, as for instance \emph{indeterminacy of identity}, which focuses on the question whether quantum particles are intrinsically indistinguishable (cf.\ \cite{French:2006aa}), and \emph{observable indeterminacy}, related with the failure of value-definiteness of quantum observables---the feature of quantum theory analyzed by Schrödinger and Dirac (cf.\ \cite{Bokulich:2014} and \cite{Calosi:2018}). Moreover, although discussions about QI are mainly concerned with the standard formulation of QM, indeterminacy is now studied in other interpretations of the quantum formalism as well, as for instance in relational quantum mechanics (cf.\ \cite{Calosi:2020}), modal interpretations (cf.\ \cite{Calosi:2022b}) and decoherence-based many-worlds interpretation (\cite{Calosi:2022}).

Against this background, many scholars underline that there are quantum theories in which indeterminacy has no place \emph{tout court}. Typical examples are frameworks implementing a clear ontology as e.g.\ Bohmian Mechanics (BM) or Ghirardi-Rimini-Weber theories (GRW). This belief seems to follow from the methodological guidelines defined by Bell’s theory of local beables (\cite{Bell:1975aa}) and further elaborated by the Primitive Ontology (PO) perspective (cf.\ \cite{Allori:2013ab}). The latter requires that in every physical theory $T$---either quantum or classical---one must (i) postulate a set of theoretical entities taken as primitive and referring to objects located in 3-dimensional space, and (ii) provide consistent dynamical laws governing their motion. It is usually argued, then, that physical phenomena and measurement results are explained and reduced to the ``histories'' of the PO.\footnote{PO theories are usually embedded within a spacetime structure which is taken to be a real substance. Relationalist reconstructions of PO theories have been proposed in  \cite{Vassallo:2015} and \cite{Vassallo:2016aa}.} Experimental outcomes are thus defined as ``functions'' of the primitive variables (cf.\ \cite{Durr:2004c}, \cite{Goldstein:2012}, \cite{Esfeld:2017}). Consequently, if every physical phenomenon or measurement result is reduced to the dynamical evolution of a well-defined primitive ontology, then there is no margin for indeterminacy.

In particular, it is a common opinion that Bohmian mechanics eliminates any sort of indefiniteness from the realm of quantum physics in virtue of its clear particle ontology.\footnote{For space reasons, in this essay I will focus exclusively on Bohmian mechanics. For a discussion of MI within GRW theories the reader may refer to \cite{Mariani:2022}.} Referring to this, in fact, Skow claims that:
\begin{quote}
[t]here are many other interpretations of quantum mechanics (Bohmian mechanics, for example, and the many Everettian interpretations) that make no use of the notion of metaphysical indeterminacy. If we reject the orthodox interpretation and accept one of those instead, then we do not have to say that there is actually any deep metaphysical indeterminacy (\cite{Skow:2010}, p.\ 856).
\end{quote}

\noindent Even more explicitly, Glick argues that:
\begin{quote}
[h]owever, none of the three most popular realist interpretations involve indeterminacy of this sort [BM, GRW and Everett's interpretation]. First, and most straightforwardly, the Bohm theory endows particles with determinate positions and momenta at all times. While it's possible that other properties (e.g., spin) may lack determinate values, position is the only fundamental feature of Bohmian particles (\cite{Glick:2017}, p.\ 2).
\end{quote}

\noindent Finally, similar views are expressed by Chen:
\begin{quote}
[w]e now have precise formulations of quantum mechanics such as Bohm's theory, GRW spontaneous collapse theory, and Everett's theory [...]. In those theories, there is no vagueness in the fundamental material ontology or fundamental dynamics (\cite{Chen:2022}).
\end{quote} 

Contrary to these claims, the aim of the present paper is twofold: firstly, I will show that even in Bohmian approaches to quantum mechanics there is room for MI. Namely, I argue that there are questions about worldly states of affairs that cannot be determinately answered by simply looking at the ontology and dynamical laws of various formulations of pilot-wave theories. I will defend this thesis discussing the case studies offered by David Bohm’s causal interpretation (\cite{Bohm:1952aa, Bohm:1952ab}) and by two generalizations of the pilot-wave approach capable of describing the phenomena of particle creation and annihilation, i.e.\ Nikoli\'c's account (\cite{Nikolic:2010aa}), and the Bell-Type Quantum Field Theory (BTQFT, \cite{Durr:2004aa}). It will be shown that these mentioned theoretical frameworks host different forms of MI. It will be furthermore argued that a new form of \emph{modal} indeterminacy, not yet discussed in the literature about QI, arises in Bohmian mechanics as formulated in \cite{Durr:2013aa}.

Secondly, it will be claimed that while the postulation of a clear ontology for a certain theory $T$ cannot a priori exclude the presence of indeterminacy in it, the latter does not necessarily threaten or undermine $T$'s ability to provide sound explanations for physical phenomena and measurement outcomes. In sum, a well-defined ontology and indeterminacy can peacefully coexist within the same framework. 

For the sake of clarity, it is worth noting that in what follows I will neither defend any specific account of QI, nor I will discuss the various perspectives on QI---for a detailed summary on this topic the reader may refer to \cite{Calosi:2021b} and \cite{Torza:2023}.

The essay is articulated as follows: Section 2 defends Lewis’ conclusion about the presence of indeterminacy within Bohm’s own causal approach from critical objections. In addition, I argue that BM hosts a modal form of indeterminacy. Section 3 analyses the presence of MI in Nikoli\'c's relativistic pilot-wave theory as well as in BTQFT. It turns out that the former entails indefiniteness concerning the properties of quantum systems, whereas the latter implies indeterminacy with respect to the future states of particles’ configurations. The philosophical implications of our discussion are discussed in Section 4, which also concludes the paper.

\clearpage
\noindent (\textbf{Partial References})
\bibliographystyle{apalike}
\bibliography{PhDthesis}
\clearpage
\end{document}